# Value-driven Analysis Framework of Service Ecosystem Evolution Mechanism

Xiao Xue, Deyu Zhou, Yaodan Guo, Zhiyong Feng, Lejun Zhang, Lin Meng

*Abstract*—With the development of cloud computing, service computing, IoT(Internet of Things) and mobile Internet, the diversity and sociality of services are increasingly apparent. To meet the customized user demands, Service Ecosystem is emerging as a complex social-technology system, which is formed with various IT services through cross-border integration. However, how to analyze and promote the evolution mechanism of service ecosystem is still a serious challenge in the field, which is of great significance to achieve the expected system evolution trends. Based on this, this paper proposes a value-driven analysis framework of service ecosystem, including value creation, value operation, value realization and value distribution. In addition, a computational experiment system is established to verify the effectiveness of the analysis framework, which stimulates the effect of different operation strategies on the value network in the service ecosystem. The result shows that our analysis framework can provide new means and ideas for the analysis of service ecosystem evolution, and can also support the design of operation strategies.

## I. Introduction

With the development of information technologies such as service science [1], cloud computing [2], Internet-ware [3] and mobile Internet, more and more enterprises and organizations encapsulate their business capabilities (e.g., resource, platform, software, business and data) into services (e.g., Web service, RESTFul service, OpenAPI and Mobile APP). These cross-organization services can be dynamically composed or coordinated through service-oriented technologies, such as Workflow, Composition/Mashup and Personalized Service. In the long-term competition and cooperation, a complicated interactive relationship and dynamical collaboration among service nodes can be formed through their self-organization mechanism. In the context of the rapid development of service-based economy [4] and software service technologies[5], service ecosystem is generated, which is featured by rapid growth, dynamic change, mutual correlation and self-adaption[6-9].

The current global market is rapidly changing and user needs are increasingly individualized. Up to this day, service ecosystem has become an important factor in the fierce global market competition. Service ecosystem is a complex socio-technical system. As shown in Fig.1, there are three main roles in service ecosystem, namely service provider, service consumer, and service operator. Service providers refer to those who are in possession of resources, and provide services to service consumers within a specific time. Service consumers refer to those who consume the resources and enjoy the services. The supply and demand matching between service providers and service consumers has been creating and producing values continuously [10,11].

This value-driven operating process determines where the ecosystem is evolving. Successful cases include Google's Android, Apple's IOS, etc.; failure cases, such as Nokia's Symbian operating system, etc. [12]. So, it is very important to clarify the value operation mechanism of service ecosystem to identify the appropriate service operation strategy. However, due to the complexity of service ecosystem, its analysis is facing the following challenges:

**Individual complexity:** In a service ecosystem, service providers are social, which increases the diversity, uncertainty and dynamics of service provision. At the same time, individuals with strong independent decision-making ability and adaptability are capable of continuous self-regulation and dynamic evolution based on environmental changes. As a result, the status and characteristics of service nodes in the ecosystem are always changing.

**Organizational complexity:** In a service ecosystem, all service providers need to benefit from collaboration on the premise of not losing their flexibility. Therefore, the frequency and degree of collaboration between services will change as their interaction relationship changes. This will affect the cost and benefits of their subsequent collaboration. Therefore, the cross-border collaborative relationship between services is unstable and always in dynamic adjustment.

**Social complexity:** In a service ecosystem, every service node has its specific function and location, and different composition forms between nodes can be utilized to fulfill complex demands. Both the source of service provision and the needs of customers are social, and this sociality has exacerbated the diversity, uncertainty, and dynamics of their supply and demand matching. Affected by this, internal system changes or external environment factors may cause unpredictable emergencies, making it hard to analyze and predict the system evolution path.

Thanks for the support provided by National Key Research and Development Program of China (No. 2017YFB1401200), National Natural Science Foundation of China (No.61972276, No. 61832014, No. 41701133). (Corresponding author: Xiao Xue)

Xiao Xue is with College of Intelligence and Computing, Tianjin University, also adjunct professor in the School of Computer Science and Technology, Henan Polytechnic University, P.R.China. (E-mail: jzxuexiao@tju.edu.cn).
Lejun Zhang is with School of Information Engineering, Yangzhou University, P.R.China. (E-mail: zhanglejun@yzu.edu.cn).

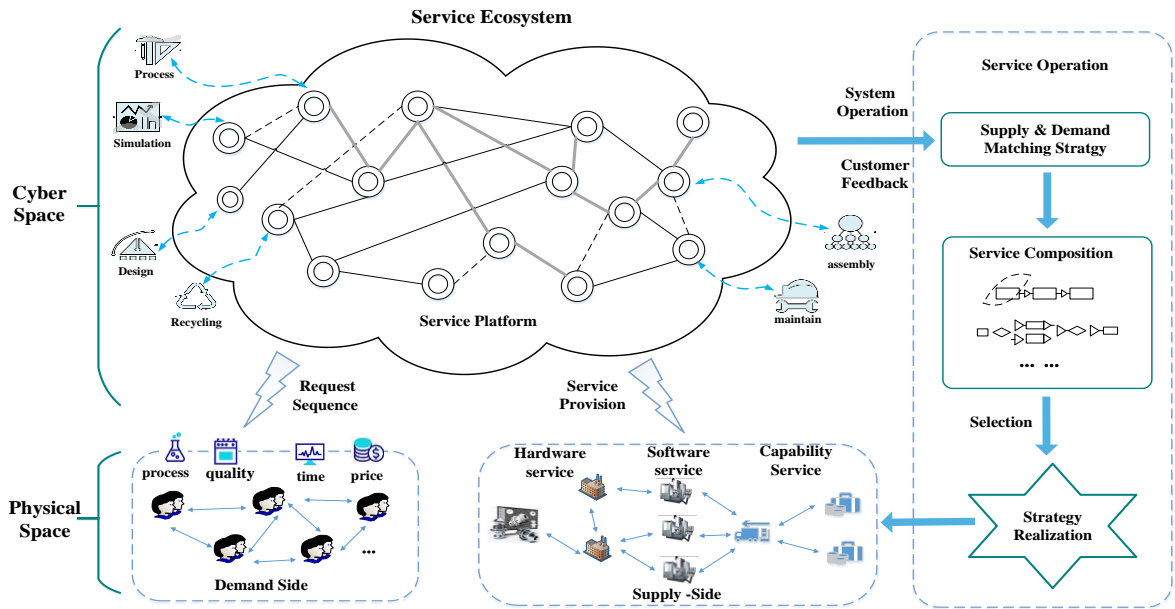

Fig.1 Operation diagram of service ecosystem

The current research mainly focuses on the analysis of static relationships in service ecosystem[13].It is difficult to reveal the complex dynamic relationship of different stakeholders in the evolution of the service ecosystem, including performance evaluation, evolution paths, evolution interventions, and so on. We need a way to conduct an in-depth analysis of the evolution mechanism of service ecosystem and reveal the operating laws behind it. The value network is the core driving force for the evolution of the service ecosystem. Inspired by the idea, we have proposed the value-driven analysis framework of service ecosystem, including value cration, value operation, value realization and value distribution. Furthermore, the computational experiment system is constructed to simulate the effect of various operation stratigies on the whole value cycle in service ecosystem.

The rest parts of this paper are organizaed as follows. Section II introduces relevant work of service ecosystem; Section III presents the value-driven analysis framework of service ecosystem evolution from the perspective of supply and demand matching; Section IV verifies the applicability of the analysis framework with various computational experiments; Section V discusses the effectiveness of our framework in practical cases; Section VI concludes the paper.

## II. BACKGROUND AND MOTIVATION

The concept of service ecosystem is originated from the ecosystem theory in ecology. Moore firstly applied the ecosystem thought in the business field and thereby proposed the concept of business ecosystem[14]. Subsequently, Vargo and Lusch proposed service-dominant logic to replace traditional commodity-dominant logic, defined the service ecosystem as a socio-technical system featured by complexity, self-evolution and autonomy [15]. The relationship between service ecosystem and value network is our reserach focus. The current research status is shown below:

*A. Analysis method of service ecosystem*

The analysis of the service ecosystem has always been the focus of academic circles, and its research is mainly divided into three categories:

**(1) Measurement and Evaluation**

To analyze service ecosystem, some researches analyze the scale, availability, complexity of services and other performance indexes (e.g., service round-trip time, throughput and utilization) by using the statistical analysis method. Masri and Mahmoud compared and analyzed the scale, availability and complexity of the obtained Internet services [16]. Zheng et al. collected 21,197 public services from the Internet and analyzed their round-trip time (RTT) and failure-rate (FT) under real Internet environment [17,18]. Cavallo et al. collected RTT of services at different time points to constitute the time sequence of QoS, and then applied the autoregressive moving average model (ARMA) to predict such time sequence [19]. Godse et al. further gave the evaluation of QoS (Quality of Service) by weighting four predicated indexes, such as RTT, throughput, accessibility and availability [20].

**(2) Evolution and Analysis**

In order to improve the undersatanding of service ecosystem, part of scholars analyze the service ecosystem from the prospective of system evolution. Alistair Barros et al. defined five main roles in Web service ecosystem, thus to discuss the provision, discovery and choreography, interaction, quality management, coordination and other key problems of services [8]. Moore pointed out that enterprises play different roles and occupy different ecological niches in service ecosystem depending on their own resources and abilities [14]. Villalba et al. designed the multi-agent-based simulation model to analyze the features of service ecosystem, including self-organization, self-adaptability and continuous evolvability, etc[21,22]. Mostafa et al. modeled each service into the independent Service Agent and defined the service composition process as the self-organization collaboration among service agents [23].

**(3) Intervention and Optimization**

In fact, the status of service ecosystem will directly decide the quality of service provision. Hence, it is very important to guide and optimize the evolutionary process of service ecosystem. Some researchers used the reinforcement learning method to deal with the dynamics and uncertainty of the Internet environment and obtain the optimized service composition [24-27]. Part of study changed the optimization problem of service network into the graph search problem, and the shortest path method was utilized to obtain the optimal solution in the service network [28, 29]. Some study started to introduce the system control concept to the study of service ecosystem. Robin Fischer, Ulrich Scholten, et al. provided a kind of feedback control-based service ecosystem frame to support the control of service provider and the management of service operator [30]. Diao proposed applying the control theory to the management of service system, and achieving the dispatching and management of service by monitoring the service quality [31].

B. *Value network of service ecosystem*

The definition of the word "Value" was first proposed by Porter in "Competitive Advantage" in 1985, that is, "the price that customers are willing to pay for goods provided by enterprises" [32]. The service ecosystem is essentially a value ecological network, which is expressed as the interaction of three heterogeneous networks, namely, the dependency network between services, the collaboration network of service participants, and the value network of service participants. The dependency network of services is a mutual reference relationship between services. The collaborative network between service participants indicates that service participants interact with each other using services as a bridge. There is an exchange of benefits among service participants in the interaction process, which leads to a value chain between service participants. The intersection of multiple value chains constitutes a value network. The current research on value networks is mainly carried out from three aspects:

**(1) Value creation.** It is about the mechanism of value generation in the ecosystem. This mechanism is often a concern for platform owners. It needs to balance the control of the coordinator (Closed tendency) with the autonomy of the participants (Open tendency). Such governance mechanisms include: autonomy priority, knowledge sharing, control right allocation, decision sharing, etc. Vargo & Lusch constructed a value co-creation model in the service system, and argued that the resource dependence between the actors resulted in service exchange, resource integration and value creation[33]. Joe Peppard et al. proposed a method of network value analysis (NVA), explaining the position of value in the network and how to create value [34].

**(2) Value realization**. It explains the mechanisms used by participants to maintain the ecosystem. Alves proposes a partnership model to define the roles and responsibilities of participants, as well as mechanisms to improve communication and collaboration within the system[35]. Such governance mechanisms include: establishment of partnership models, definition of roles and responsibilities, conflict and risk management, etc. In [36], the value realization process among stakeholders is demonstrated, including direct value exchange (that is, direct payment for services provided and used) or indirect value exchange (that is, revenue generated through advertising) . Touliou et al. [37] explained the value proposition of stakeholders from a business perspective and gave a description paradigm of value proposition. Haile et al. [38] used covariance analysis to evaluate the value of software service platforms, considering the impact of different roles on the realization of system value, including QoS (Quality of Service), service developers, service platforms, users, and service prices.

**(3) Value distribution.** The service ecosystem needs to attract and retain participants through value distribution. Such governance mechanisms include: income distribution models, incentive mechanisms, investment and cost sharing, and so on. Pant et al. proposed a goal-oriented value analysis framework, using supply chain dependency analysis and cash flow analysis to judge the operation of the ecosystem[39]. Cong P et al. proposed a dynamic pricing model based on the perceived value of users, which maximized the value of cloud service providers by capturing the supply-demand relationship in the cloud service market [40]. Based on the premise of meeting the profit of service providers, the value-driven service system design methods was developed to improve customer satisfaction [41].

In order to make the service ecosystem evolve in the expected direction, we hope to explore in depth how service operation strategies affect the value network and thus drive the evolution of the entire ecosystem. However, the existing research is fragmented and unsystematic, and cannot effectively reveal the relationship between the value network and the service ecosystem. Based on this, we propose a value-driven analysis framework that can reveal the relationship between value network operation stratgegy and service ecosystems evolution mechanism.

III. VLAUE-DRIVEN ANALYSIS FRAMEWORK OF SERVICE ECOSYSTEM

The service ecosystem is a complex socio-technical system with self-organizing and co-evolving characteristics. In the service ecosystem, there are three important roles: service provider, service consumer, and service operator. The value cycle among the three roles drives the operation and evolution of the entire system. In order to clarify the evolution mechanism of the service ecosystem, this section will analyze the operation of the value network and its feedback impact.

A. *Value Creation of service ecosystem*

The evolution of the service ecosystem depends on the dynamic formation of value networks between service providers and service consumers. The demands of service consumers is the driving force of the entire value network operation, and determines the final output of the service ecosystem. The collaboration network of service providers is the executor of the value network and determines the operating costs of the service ecosystem. Different service operation strategies will lead to large differences in the output or cost of the value network, which will affect the final evolution of the service ecosystem.

In the process of meeting the needs of service consumers, the value chain between the service provider and its partners is formed. When the demand is completed, the value chain realizes value creation, that is, the service consumer transfers profits to the service providers. The diverse composition of different value chains will form a value network, which can provide a basis for creating more value together.

As shown in Fig.2, the operation of the service ecosystem is a value-driven cyclical feedback, which is composed of four steps: value creation, value operation, value realization and value distribution. The value network is a virtual network with a topological equivalence mapping relationship with the service network, which is positively generated by service providers and service consumers through supply and demand matching; and the evolution of the value network will in turn affect the dynamic adjustment of service providers and service consumers.

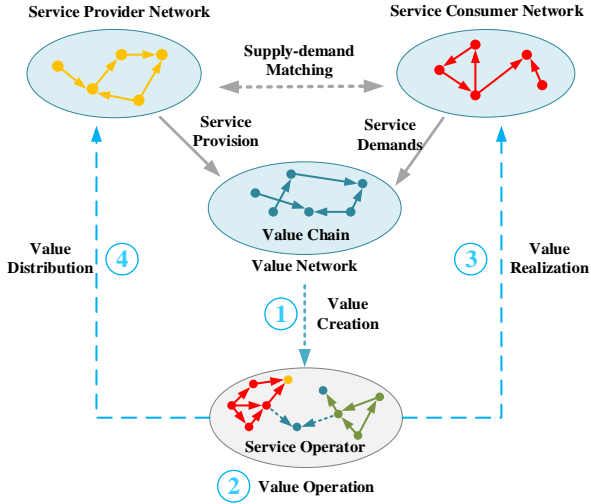

Fig.2 Operation diagram of service ecosystem

The service operator is the link between service provider and service consumer, and its operation strategy directly affects the generation and operation of the value network. In the service ecosystem, the source of service provision and servcie consumers are social. This sociality exacerbates the diversity, uncertainty, and dynamics of service provision and service demands. In order to promote the matching between supply and demand, the service operation strategy needs to continuously adjust the resource allocation according to the current state of supply and demand matching. This kind of decision depends on the corresponding decision knowledge and utility function.

The supply and demand matching can be met by multiple value chains. In order to achieve the best matching between supply side and demand side, it is necessary to determine which value chain can maximize the value creation. Here, the behavior mechanism of the service operator is given as follows:

$$S_I \times S_C \times K_P \to \Delta \quad (1)$$

Among them, $S_I$ represents the state space of service resources and subsystems of the supply network; $S_C$ represents the state space of the demand network; $K_P$ is the strategic knowledge set, including the organizational model of network collaboration, task assignment rules, revenue distribution rules, etc.; $\Delta$ is the final selected service operation strategy. Its evaluation functions is as follows:

$$MAXIMISE(\ value,\ configPara,\ mc,\ min,\ max,\ step) \quad (2)$$

Among them, *value* represents the utility function of service operation strategy; *configPara* represents the configuration parameters of the service operation strategy; *mc* represents the Markov chain model of the demand sequence, etc., which can support the setting of *configPara*; *step* represents the size of the parameter adjustment, its value is between *max* and *min*. According to the concept of value engineering[42], the calculation function of variable *value* is defined as follows:

$$Value = \frac{Outcome}{Cost} \quad (3)$$

Here, *Outcome* indicates the overall benefits of the service ecosystem in the supply-demand matching process, including the profits earned on the supply side and the user satisfaction on the demand side. *Cost* represents the total cost of delivering services throughout its life cycle. *Value* represents the ratio of *Outcome* to *Cost*.

According to formula 3, there are two main directions to create more value: increasing output or reducing costs. The strategy to increase output is mainly to promote technological innovation, such as service meshup, traffic diversion and others. The strategy to reduce costs is mainly to promote healthy competition among service providers, such as scoring strategies, charging strategies, and so on. According to whether the calculation result of the *Value* variable is much greater than 1.0, the evolution of service ecosystem has two typical states:

(1) **Value explosion state**: In the phase, service providers compete based on technological innovation, which can create high value-added products and services. Overtime, as technology spreads and similar services emerge, technological innovation decreases while process innovation increases.
(2) **Value capturing state**: In the phase, service providers compete on cost rather than technology innovation. With the survival of the fittest, economies of scale will lead to the exit of smaller participants and increase barriers to entry. In the end, the ecosystem gradually entered a steady state until a new technological breakthrough appeared.

*B. Value operation of service operator*

The purpose of the service strategy is to influence the value network by adjusting the relationship between supply side and demand side. The performance of different service strategies varies widely. Therefore, the design of service strategy needs to consider many factors, including how the implementation time and length of the strategy will affect the results; what changes (e.g. smooth change or drastic change) does the system show when the strategy changes; how to balance between the intensity of strategy implementation and cost control. To create value as much as possible, we applies a complex networked systems lens to identify and analyze the performance of two kinds of service strategies shaping the service ecosystem: convergence and coopetition.

(1) **Convergence: the strategy to increase outcome**

With the diversification and complexity of user needs, single-functional services often fail to meet user needs. Service convergence refers to a transformation process that blurs boundaries by unifying value propositions, technologies, or markets. It can meet the diverse needs of users through service innovation. According to the depth of service convergence, the

convergence strategy can be divided into the following three categories:

**Information convergence.** It is the combination of information bases, which canerode the boundaries of those isolate services. Internet traffic guidance is a typical information convergence strategy, which can help distinct online services to cite each other. In this way, the current service can attract some user traffic from other services. The current service needs to pay the portal service some fees for the increase in the number of users.

**Capability convergence.** It is defined as the combination of previously distinct services into a common service. Capability convergence is generally motivated by the potential of combining own capability with external one, thereby leading to new value-creating opportunities and innovative service offerings. For example, the bundling of Location service and Map services create Navigation services, and the bundling of Online retail services and Offline supermarket services create a New Retailer solution.

**Domain convergence.** When applications from distinct domains are combined, they infringe on existing value-creating territories of underlying domains and industries. Domain convergence often leads to a new crossover service that widens markets, lowers barriers to entry and increases competition. Moreover, domain convergence can lead to reconfiguration of the value chain through the addition or elimination of activities, consolidation through mergers and acquisitions, etc. It is particularly prevalent in the Internet context. e.g. "Internet + traditional retail", "Internet + finance", "Internet+ tourism", etc.

**(2) Coopetition: the strategy to reduce cost**

In today's dynamic business environment, srevice nodes have to compete and cooperate at the same time in order to grow and survive. The cooperative aspect of coopetition refers to the collective use of shared resources to pursue common interests; the competitive aspect refers to the use of shared resources to make private gains in an attempt to outperform partners. According to many economic literatures, the profit sharing ratio between different value links is very important to the development of the coopetition relations [43].

On the one hand, coopetition can accelerates R&D efforts, significantly reduces costs, diversifies the portfolio of products or services, and drives higher links of consumer satisfaction. On the other hand, we must take these negative and unintended effects into account, such as the conflicts between individual interests and overall collaboration goal. Here, we design three profit sharing strategies, which are taken as the experiment objects. The related details are given as follows.

**High-fair strategy:** It emphasizes the overall competitiveness and stability of the ecosystem. The profit sharing ratio between different value links is relatively balanced. The service nodes in different links can obtain enough profits to survive. The overall synergy of the service ecosystem can be fully utilized.

**Moderate-fair strategy:** It emphasizes the pursuit of the balance on the premise of maintaining the interests of the leading links. The profit sharing ratio between different value links is not so balanced. Most service nodes can only survive and the overall effectiveness of the ecosystem cannot be fully utilized.

**Low-fair strategy:** It emphasizes the benefits of the dominant link. The profit sharing ratio is very uneven. Most of the benefits are captured in the dominant link, and other links will struggle to survive by the ecosystem.Thus, the overall service quality continues to decline, and customers also pay unnecessary costs such as time and prices, which may even cause the entire ecosystem to be unsustainable.

*C. Value realization of service consumers*

The demand side is the source of value for the entire ecosystem and the key of driving the value network. Service consumers can choose available services and resources according to the current situation and their own preferences, thereby fulfilling their various needs. The operating mechanism of the service consumer can be composed of a set of interrelated behaviors and decisions. The formula is as follows:

$$R_k = (\{S_i : r(S_i) = R_k\}, \{D_i : r(D_i) = R_k\}, \{M_i : r(M_i) = R_k\}) \quad (4)$$

Here, $S_i$ represents the demand list associated with the role, that is, the products/services required by the role; $D_i$ represents the decision mechanism associated with the role, that is, the criteria by which the role selects the service provider, such as service category, quality interval, price interval, and delivery location etc.; $M_i$ represents the metric associated with the role, which is used to measure the benefits of selecting the service and provide a basis for adjusting the selection.

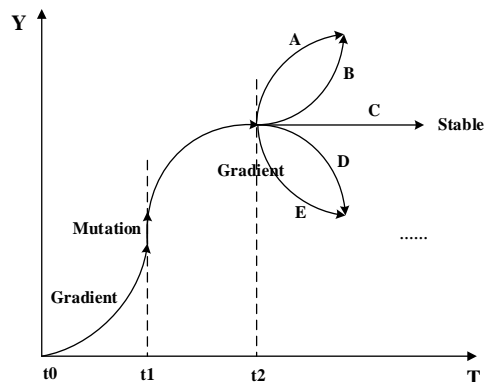

Fig.3 Value realization process of service ecosystem

The value realization of the service ecosystem is mainly to measure whether the interests of the demand side have been met. Based on the ecological theory and population growth theory, Fig.3 shows the value realization process of service ecosystem: the *Y* axis represents value, and the *T* axis represents time. The change of the value curve of the service ecosystem is divided into three trends.

**Upward trend**: Types *A* and *B* indicate that the ecosystem develops in a new direction after finding a new value growth point, and the growth rate curve may grow rapidly or tend to grow steadily. In the case, the positive inter-group network externality between demand and supply will play an important role. Service consumers can be better off in the ecosystem when the number of service providers increases, and vice versa. If there are enough services in the service ecosystem, the lower the cost for users to find suitable services, the higher the corresponding value output. The individual choice of service consumers may continue to attract new users, resulting in an explosive growth in the number of users.

**Downward trend**: Types *D* and *E* indicate that after a period of development, the ecosystem gradually declines due to competition from other new systems. Type *D* indicates rapid decline, and type *E* belongs to slow decline. When the ecosystem has reached a certain link, the negative intra-group network externality among users begins to emerge. It means that the value realization of an individual user can be reduced when more service consumers join the same side. This will lead to an increase in the time cost of selecting and waiting for services, resulting in less self-satisfaction. At this time, service consumers will continue to withdraw, which may affect the surrounding individuals and cause the overall number of users to shrink.

**Steady trend**: Type *C* indicates that the overall trend of service consumers has not changed much, and the value link of the system is in a relatively stable state.

User satisfaction is neither easy to measure objectively nor easy to obtain. Therefore, in order to evaluate the impact of the value network on service consumers, the following indirect indicators are generally used: ① The number of active users of each service. According to the change of this indicator, we can see the overall trend of demand change, and predict the possible inflection point of value. ② Gini coefficient for similar services. Here, the number of users calling the service is analogized to the income of residents. The calculated Gini coefficient can be used to measure the differences between services. ③ User traffic guidance ratio among services. This indicator can be used to evaluate the synergy efficiency between services. However, it is generally difficult to obtain relevant data.

*D. Value distribution of service providers*

Service providers serve as the active service entities in service ecosystem, i.e. the active service nodes in the value network. They have some typical characteristics, such as interconnection rather than isolation, autonomy rather than obedience, etc. Each service node not only needs to know its own division of labor, but also needs to choose the right partner across the organization. As shown in Fig.4, through various forms of negotiation processes among multiple service nodes, a loosely coupled, dynamic, and common goal-oriented value chain is formed to meet the increasingly complex and diverse customer needs. Organizational forms include the collaboration within the alliance, the collaboration between alliances, and the collaboration across alliance boundaries.

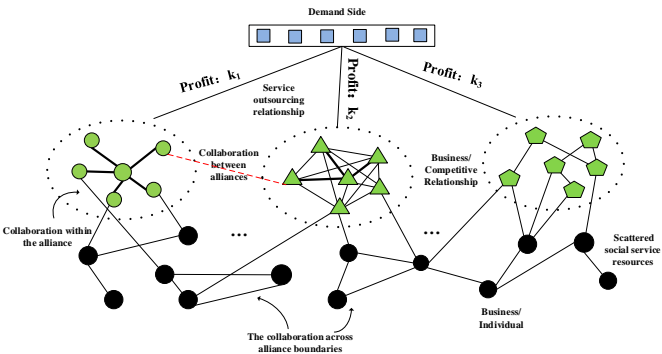

Fig.4 Different types of aggregation relationships of service nodes

The value distribution among service providers are key factors driving the evolution of supply network. In the intense competition among service nodes, those nodes that are not competitive are likely to be eliminated. In order to survive in the ecosystem, service nodes must improve their decision-making and behavioral skills through a variety of learning methods. The evolution process of service provider is the result of the combined effects of individual evolution, organizational evolution and social evolution. Based on the work in [44], the SLE framework is used to describe the characteristics of the service providers.

**(1) Individual evolution layer**

It is used to depict the independent evolution of individual service provider in the real world. According to the rule of survival of the fittest, each individual node needs to continuously improve its own ability in order to survive in the fierce market competition. Service providers are generally described as follows:

$$Service\ Provider = \langle R, S_t, E_t, V_t, Y_t, N \rangle \quad (5)$$

Where, $R$ is the role of the service provider in the system, which is used to define the basic value of the service it provides; $S_t$ is the capability property of service provider, including all tangible or intangible resources required for service provision, which can change with time; $E_t$ is the perception capability of service provider, including its perception channels and scope; $V_t$ is a collection of behavior that service provider can perform, including its spontaneous behavior and all the actions triggered by external events; $Y_t$ is the operation strategy that the service provider makes decision autonomously; $U_t$ represents the adaptive mechanism of updating its own operation strategy, such as reinforcement learning, observational learning, imitation learning, and so on.

**(2) Organizational evolution layer**

It is mainly used to depict the cooperation between service nodes to enhance the competitiveness. In the real world, market competition has evolved from the competition between single nodes to the competition between groups. However, there is a conflict between maximizing individual interests and maximizing overall interests. Therefore, the cooperation relationship between service nodes often needs to be adjusted according to the actual situation. Whether the collaboration between different service nodes can be formed depends mainly on the following value distribution relationships:

$$\begin{cases} T\_Out_M \geq T\_Out_M^* \\ T\_Out_R \geq T\_Out_R^* \end{cases} \quad (6)$$

Here, $T\_Out_M$ and $T\_Out_R$ indicate the benefits obtained by the two service nodes when they choose the collective benefit maximization strategy. $T\_Out_M^*$, $T\_Out_R^*$ represents the benefits obtained by the two service nodes when they choose the individual benefit maximization strategy. The service nodes will be in a long-term game relationship. Only when the above formula is satisfied, the two service nodes will tend to participate in the collaboration.

**(3) Social evolution layer**

It is mainly used to depict the impact of elite culture on individual evolution in society. In the real world, some elites with excellent knowledge will gradually emerge from the group because of their excellent performance. Then, their knowledge can be extracted into culture, and it can affect the individual evolution at the micro link. For example, the operation mode of service ecosystem (fair mode or unfair mode) can accelerate or hinder the development of many single nodes in different scenarios. Whether the collaboration between different service nodes can be sustained depends mainly on the following value distribution relationships:

$$\frac{Out_M}{Cost_M} = \frac{Out_R}{Cost_R} \rightarrow V_M \approx V_R \quad (7)$$

Here, $Out_M$ and $Out_R$ indicate the benefits obtained by the two service nodes from the collaboration; $Cost_M$ and $Cost_R$ represent the costs of the two service nodes in the collaboration; $V_M$ and $V_R$ indicate the value of the two service nodes. The benefits they receive should increase as costs and risks increase. When service providers evaluate their collaborative relationships, they must compare their input-output ratios with the average level of society as a whole. The degree of fairness will directly affect the enthusiasm of the nodes to participate in collaboration. Therefore, the distribution of value must consider both the importance of the nodes to the cooperation and the resources invested by the nodes.

## IV. COMPUTATIONAL EXPERIMENTS OF SERVICE ECOSYSTEM EVOLUTION

In this section, various experiment scenarios are designed to compare the effect of different service strategies on the value-driven evolution of service ecosystem. The experiment results will be used to verify the validity of the value-driven analysis framework.

### A. Initialization of Computational Experiment

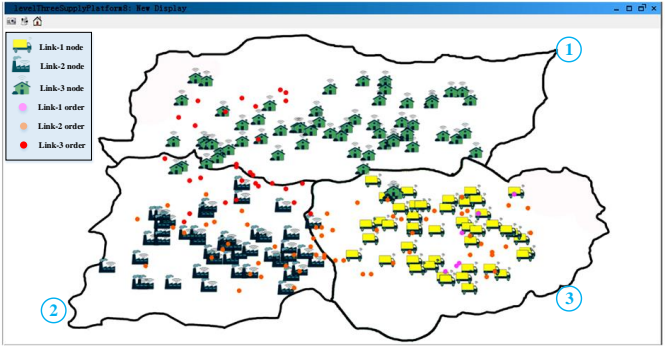

Fig.5 The operation scenario of computational experiment system

As shown in Fig.5, the entire scene is divided into 3 areas, which are occupied by different types of service nodes (first-link nodes, second-link nodes, and third-link nodes). These service nodes are active and dynamic, serving as the active behavior entity in system environment, which are all represented by different symbols. All service nodes search their own specific orders in the environment and consume certain capital in the searching process. After acquiring orders, they will earn corresponding profits and increase in capital value. When their capital reaches the reproduction threshold, genetic evolution is conducted to produce new childs of the same kind. When their capital is smaller than their death threshold, they die and disappear. Three types of nodes are randomly scattered in specific areas, forming a virtual "food chain" ecosystem.

The emergence of new orders will follow some market rules. If one order is processed by some service node in time, it can be converted into this node's profit; otherwise, the order will be lost. The survival of the fittest among service nodes are key factors driving the evolution of service ecosystem. In the intense competition among service nodes, the service nodes that are not competitive are likely to be eliminated. In order to survive in the ecosystem, service nodes must improve their decision-making and behavioral skills through a variety of learning methods. In this experiment, the adaptive learning rule of the service node is to continuously searching the order-rich area by means of observational learning and imitation learning.

In the computational experiment environment, various operation stratigies of service ecosystem can be evaluated, including some pressure test and boundary test. By observing the evolution phenomena of ecosystem in experiment system, it is possible to intuitively find the appropriate operation strategies. Here, we give the initialization part of the computational experiment, including the research objectives, parameter settings and evaluation criteria. The related details are given as follows.

**(1) Research objectives of the experiment**

The first case focuses on the operating costs of the value network. It is used to evaluate the impact of different service coopetition strategies on the evolution of the value network. The second case focuses on the effective outcome of the value network. It is used to evaluate the impact of different service convergence strategies on the evolution of value network.

**(2) Environment settings of the experiment**

The experimental environment is 200 cells in length and 90 cells in width, and different types of service nodes are randomly distributed in their respective areas. The settings of basic experimental parameters are shown in Table 1. To facilitate the comparison, all the related parameters are scaled to the same range on the basis of practical data [45,46].

TABLE 1 PARAMETER SETTINGS OF TWO EXPERIMENTS

| System variable | Experiment setting | Remark | |
|---|---|---|---|
| | | Case 1 | Case 2 |
| The generation rule and distribution rule of orders | The market trends are represented by the function $Y = N + M * \sin(t)$. In the experiment, the reference value of the order quantity $N$ is set to 100, and the fluctuation value $M$ is set to 5. Each order contains three parts of profit (k1, k2, k3), which are processed by three types of service nodes. | Orders are initially distributed in area 1. After the order is processed by the link-1 node, it will appear in the adjacent position of the next-link area. The completion of this order requires the cooperation of three types of nodes. If one service node does not complete the corresponding task, other service nodes will not be able to earn the profit of the order. | Orders are randomly distributed in three areas with fixed locations. The completion of one order does not require the cooperation of three types of nodes. Each type of service nodes is responsible for processing a certain part of the order and obtaining corresponding profits. It has nothing to do with whether the other two types of nodes have processed the order |
| Distribution rules of service nodes | Initial settings of order distribution: Link-1 nodes are distributed in area 1; Link-2 nodes are distributed in area 2; Link-3 nodes are distributed in area 3. | Different types of nodes can only be active in their own regions and cannot move across regions. But they can move towards order-rich areas to increase their chances of survival. | When there is a cooperative relationship between different regions, their service nodes can freely move and capture orders across areas. This is similar to user traffic guidance between different apps. |

| | | Service coopetition strategy mainly refers to the profit sharing ratio between upstream and downstream nodes. In the three strategies, the profit sharing ratios of the three areas are set as follows:<br>**Low fair strategy** : (1:2:7);<br>**Moderate fair strategy** : (2:2:6);<br>**High fair strategy** : (3:3:4). | Service convergence strategy mainly refers to the user traffic guidance mechanism between different regions.<br>**Non-convergence strategy:** The node searches for orders only in its own area.<br>**Partial-convergence strategy**: The nodes in area 1 and 2 can share orders.<br>**Full-convergence strategy**: Nodes can search for orders in three areas. |
|---|---|---|---|
| Service Strategy | Case 1 is mainly used to evaluate the impact of three different service coopetition strategies.<br>Case 2 is mainly used to evaluate the impact of three different service convergence strategies. | | |
| Initial number of service nodes | Area 1: Area 2: Area 3 = 50 : 50 : 50 . | The initial number of nodes in different regions is the same. Differences in service strategies are the only factors that affect experimental result. | |
| Death threshold | The capital value is 20. | When the capital value of the service node is lower than the death threshold, the node will die. | |
| Reproductive threshold | The capital value is 300. | This variable represents the capital threshold at which the service node produces child nodes. | |
| Speed | Bounded random within the range of [1, 4]. | This variable represents the ability of the service node to capture orders. | |
| Vision range | Bounded random within the range of [3, 9]. | This variable represents the ability of the service node to search orders. | |
| Distance cost | Y=k*x(x>0, x indicates the distance moved. k=0.8 ) | This variable represents the cost consumed by the node when searching for orders. | |
| Operation cost | Bounded random within the range of [1, 5]. | This variable represents the fixed cost consumption of the node in each cycle. | |
| Initial capital value | Bounded random within the range of [100,120]. | This variable represents the initial capital value of each node. | |
| Process capability | Bounded random within the range of [2,10]. | This variable represents the order processing capacity of each node. | |

**(3) Evaluation index of the experiment**

The experiment scenario can intuitively show the evolution of the service ecosystem. In Case 1, we used three indicators to evaluate the impact of different competition strategies on the cost of value network, including the number of service nodes alive, the average cost of nodes and total value of all nodes. In Case 2, we used three indicators to evaluate the impact of different convergence strategies on the outcome of value network, including the number of capturing orders, average profit of nodes, and total value of all nodes.

*B. Case 1: The impact of service coopetition strategy on value network*

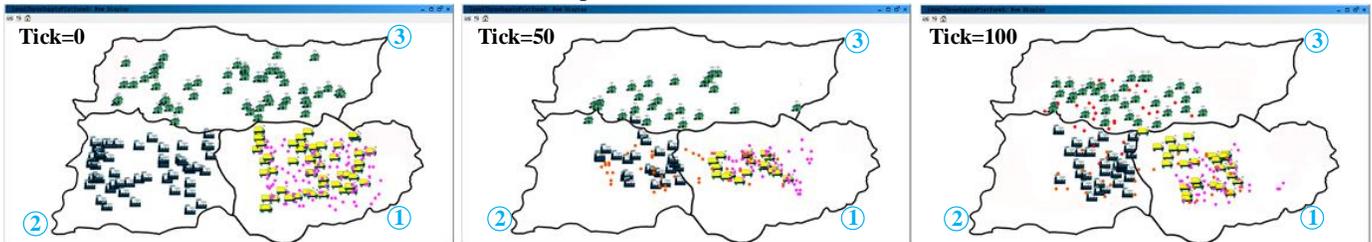

(a) The evolution trend of service ecosystem when the initial profit sharing ratio is 1: 2: 7

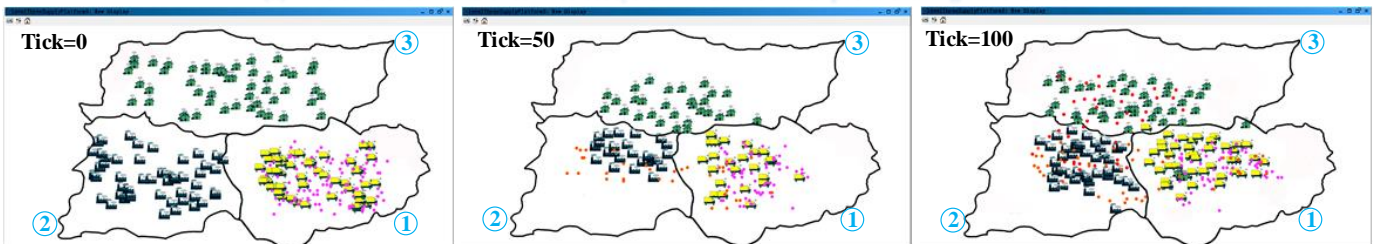

(b) The evolution trend of service ecosystem when the initial profit sharing ratio is 2: 2: 6

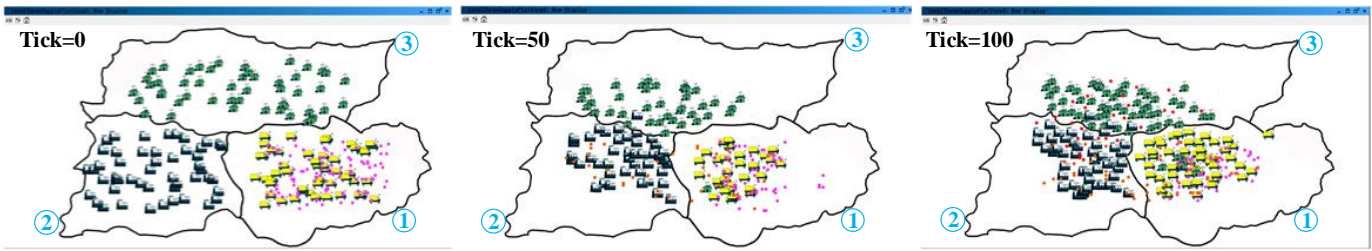

(c) The evolution trend of service ecosystem when the initial profit sharing ratio is 3: 3: 4

Fig.6 Evolution of the service ecosystem when adopting three initial profit sharing ratios

The experiment is mainly used to evaluate the impact of coopetition strategy on the evolution of service ecosystem. In the experiment, the initial profit sharing ratios of the linkl-1 nodes, link-s nodes, and link-3 nodes are set to 1:2:7, 2:2:6, 3:3:4, respectively. The top, middle, and bottom row in Fig.6 represent the evolution scenario of the service ecosystem when the three competing strategies are adopted respectively. The following conclusions can be clearly observed:

(1) When the coopetition strategy is very unfair (Fig.6-a, initial profit sharing rate is 1: 2: 7), the first-link nodes earn the least profit, which leads to their low enthusiasm for orders. As a result, a large number of orders are invalidated. Affected by the link-1 nodes, the other two types of nodes cannot obtain enough orders to survive. At the 100th cycle, the number of service nodes in all three regions was quite small.

(2) When the coopetition strategy is unfair (Fig.6-b, the initial profit sharing rate is 2: 2: 6), the profit that the link-1 nodes obtain from the order can survive, thereby increasing the enthusiasm of the link-1 nodes for capturing the order. As a result, the overall completion rate of orders has improved a lot. Therefore, the number of orders captured by the link-2 and link-3 nodes is large. Finally, the survival rate of service nodes is significantly higher than that of scenario 1.

(3) When the coopetition strategy is fair (Fig.6-c, the initial profit sharing rate is 3: 3: 4), each link of nodes has a high motivation to search for orders. The overall completion rate of orders has been greatly improved. As a result, the mortality of upstream nodes is effectively reduced, and the survival rates of nodes in the three regions are improved. The results prove that this fair coopetition strategy performs better than the other two strategies.

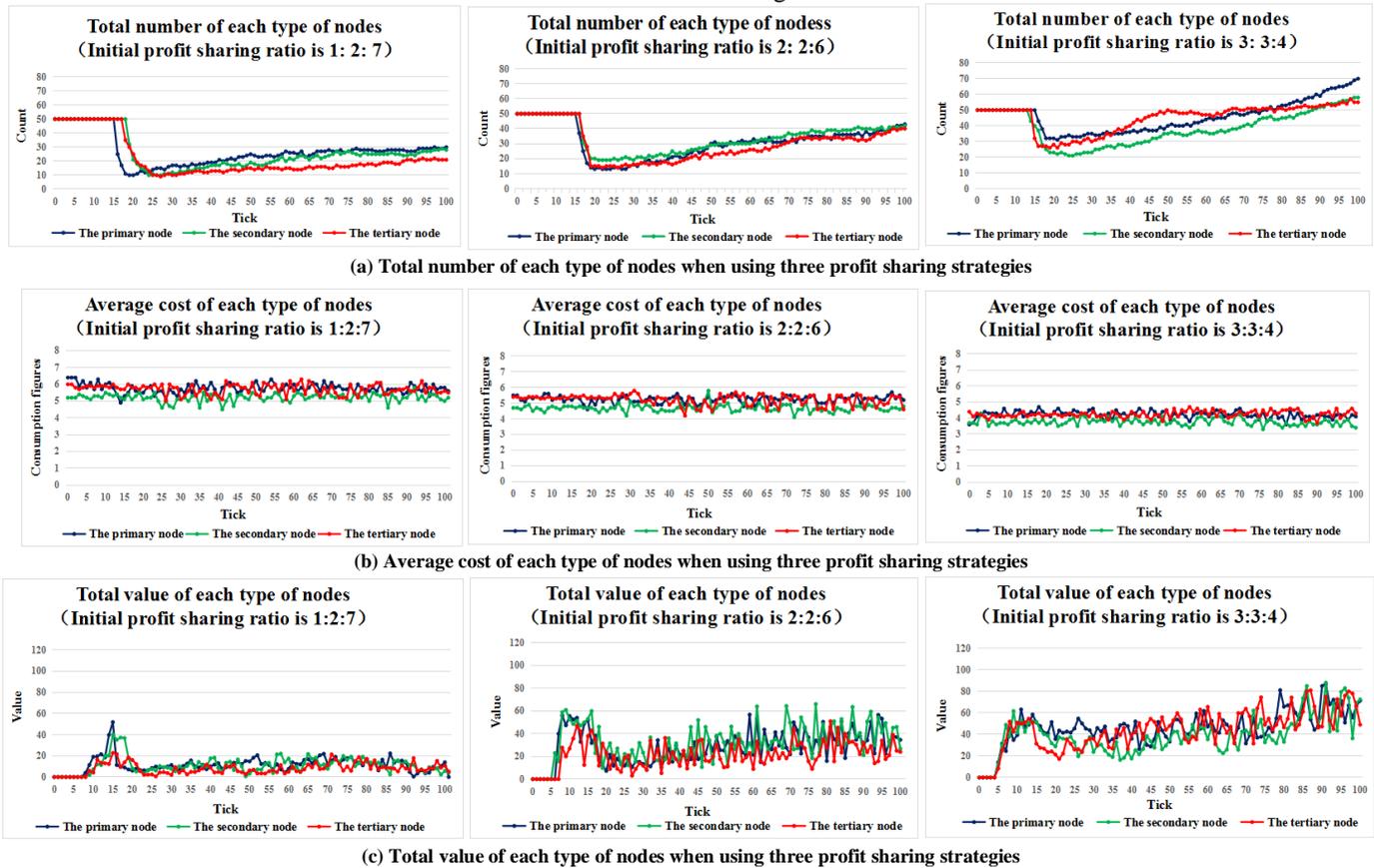

(a) Total number of each type of nodes when using three profit sharing strategies

(b) Average cost of each type of nodes when using three profit sharing strategies

(c) Total value of each type of nodes when using three profit sharing strategies

Fig.7 Comparison of three performance indicators when adopting three initial profit sharing ratios

Fig.7 gives a comparative analysis of the performance indicators of the three coopetition strategies in Case 1. The details are shown as follows:

(1) Number of nodes (Fig.7-a): ① If the coopetition strategy is very unfair, many nodes will die or exit. In the first few cycles, the number of nodes dropped sharply, and the overall recovery rate was slow. ② If the coopetition strategy is relatively fair, nodes at all links can have a better chance of survival. The overall recovery speed of the number of nodes is relatively fast. ③ If the coopetition strategy is fair, the number of nodes recovered quickly. What's more, the number of surviving nodes was significantly higher than the first two strategies.

(2) Average cost of nodes (Fig.7-b): During this evolution, orders in the central area where the three areas intersect are relatively abundant. Nodes near the central area have a higher chance of survival, while nodes far away from the central area are more likely to die. ① When the coopetition strategy is very unfair, the aggregation phenomenon is mostly a passive result caused by the elimination of nodes in remote areas. Therefore, the degree of aggregation is low and the cost of searching orders is high. ② When the coopetition strategy is relatively fair, the orders in the central area begin to increase. As a result, the degree of aggregation of service nodes has been improved, and the cost of searching orders has been reduced to a certain extent. ③ When the coopetition strategy is fair, the orders in the central area are very rich, resulting in a significant increase in the degree of node aggregation and a significant reduction in the costs of searching orders.

(3) Total value of nodes (Fig.7-c): ① When the coopetition strategy is very unfair, the overall capital scale of the three areas is greatly damaged. Due to high costs and low profits, the final value is also the lowest. ② When the coopetition strategy is relatively fair, the total value and growth rate of the three area have increased slightly. ③ When the coopetition strategy is fair, the overall capital scale of both upstream and downstream nodes will benefit greatly. Because the cost is low and the profit is high, the final value is also the highest.

*C. Case 2: The impact of service convergence strategy on value network*

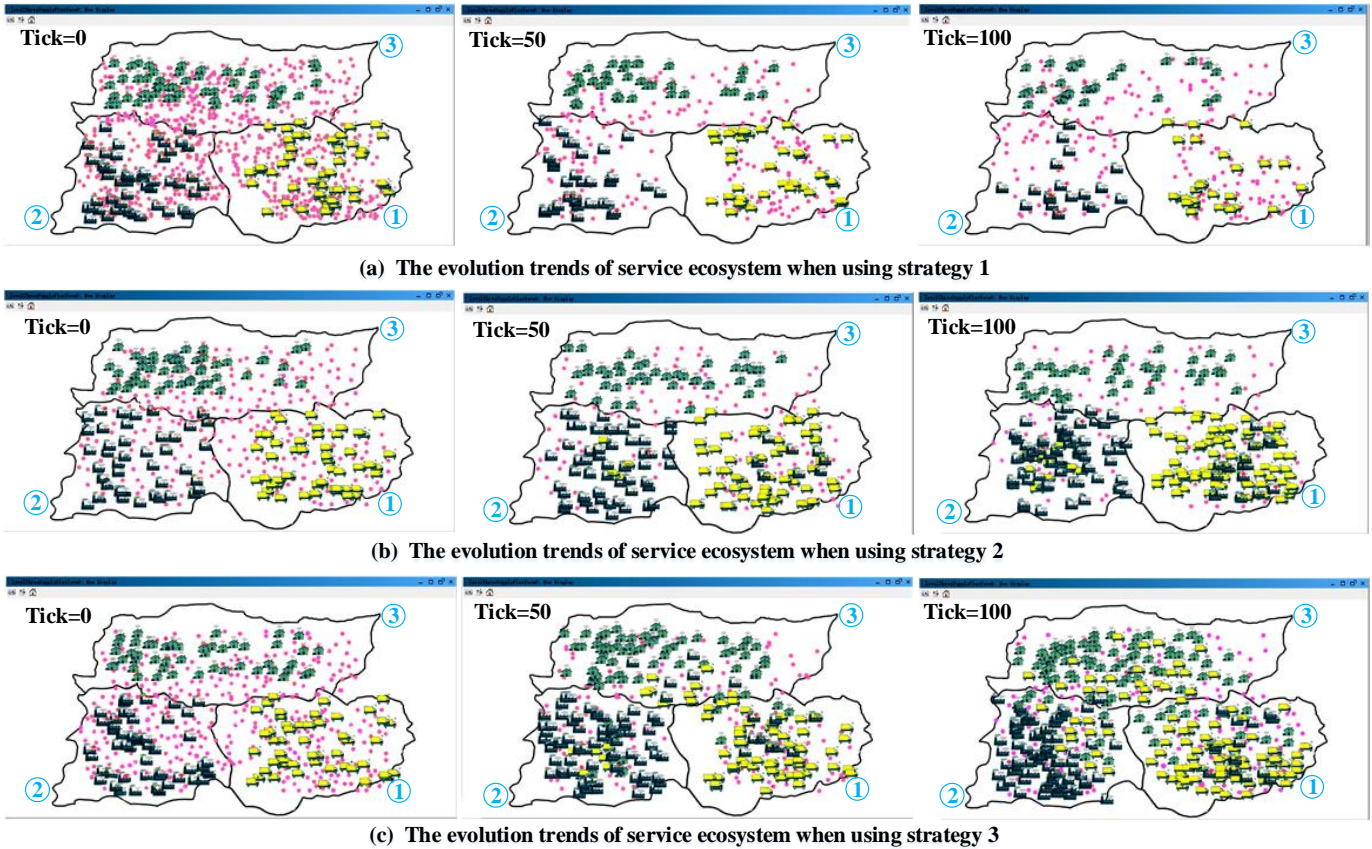

(a) The evolution trends of service ecosystem when using strategy 1

(b) The evolution trends of service ecosystem when using strategy 2

(c) The evolution trends of service ecosystem when using strategy 3

Fig.8 Evolution of the service ecosystem when adopting three convergence strategies

The experiment is mainly used to evaluate the impact of convergence strategies on the evolution of service ecosystem. In the experiment, three types of convergence strategy are adopted, including non-convergence strategy (the service node only searches for orders in its own area), partial convergence strategy (the service nodes in area 1 and area 2 can share orders), and full convergence strategy (the service nodes in three areas can share orders). The top, middle, and bottom row of Fig.8 represent the evolution scenario of the service ecosystem when three different strategies are adopted. The following results can be clearly observed:

(1) When strategy 1 (non-convergence) is used (Fig.8-a), there is no collaboration between different areas, and various service nodes are only active in their own areas. This makes many nodes have limited orders available, and cannot meet their survival needs. With continuous evolution, the number of nodes gradually decreases.

(2) When strategy 2 (partial convergence) is used (Fig.8-b), service nodes in area 1 and area 2 can search for orders freely in these two areas. This is equivalent to user traffic guidance between the two types of services, which greatly increases the probability of service nodes capturing orders. Compared with the first strategy, the probability of capturing orders in these two areas is greatly improved. However, the nodes in area 3 still cannot get enough orders to survive because they have not participated in the convergence. Finally, the nodes in area-3 are still very sparse.

(3) When strategy 3 (full convergence) is used (Fig.8-c), all nodes can search for orders in these three areas. This is equivalent to mutual user traffic guidance between the three types of services, which greatly increases the probability of obtaining orders. Compared with the other two strategies, this strategy can greatly improve the survival probability of service nodes. The overall density of nodes in the three regions is significantly higher than that of the other two strategies.

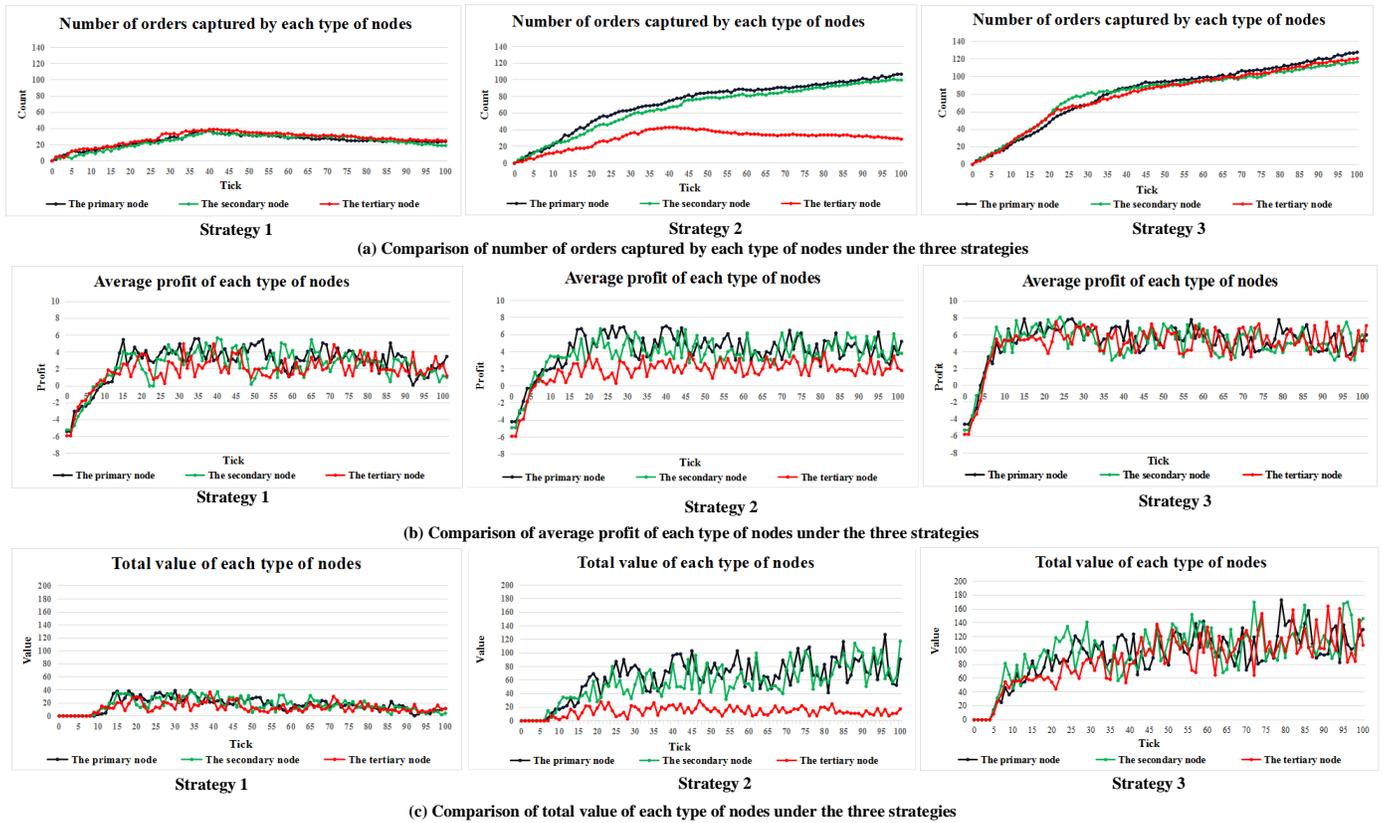

Fig.9 Comparison of three performance indicators when adopting three convergence strategies

Fig.9 gives a comparative analysis of the performance indicators of the three convergence strategies in Case 2. The details are shown as follows:

(1) Number of orders (Fig.9-a): ① When using strategy 1 (non-convergence), the three areas are isolated from each other. The number of service nodes continues to decrease, and the number of orders that can be processed is also very small. ② When using strategy 2 (partial convergence), service nodes in area 1 and area 2 increase their probability of capturing orders. The number of orders available to service nodes in these two regions has increased significantly. ③ When strategy 3 (full convergence) is used, the nodes in the three regions greatly increases the probability of the ecosystem capturing orders. Compared with the other two strategies, the number of orders processed by the three types of nodes has reached the maximum.

(2) Average profit of nodes (Fig.9-b): ① When using strategy 1 (non-convergence), the service node is affected by fixed costs and searching costs at the initial stage, and its profit is negative. In the middle stage, as the number of captured orders increases, profits increase slightly. In the later period, the profits of service nodes are in a low-level turbulence. ② When using strategy 2 (partial convergence), the profit of service nodes in area 1 and area 2 increases significantly, while the profit of service nodes in area 3 does not change much. ③ When using strategy 3 (full convergence), the profits of the three types of nodes all increase. The overall profit situation is better than the other two strategies.

(3) Total value of nodes (Fig.9-c): ① When using strategy 1 (non-convergence), the cost of capturing orders in each area is relatively high and the profit is small. It makes the value of nodes lower and tends to decrease. ② When using strategy 2 (partial convergence), service nodes in area 1 and area 2 increases the probability of catching orders. As a result, these two regions have higher profits and lower costs, so they have higher value. ③ When using strategy 3 (full convergence), the profit levels of the three areas have been greatly improved, while the costs have been continuously reduced. This makes the total value of the three types of nodes significantly better than that of the other two strategies.

V. DISCUSSION

This section will demonstrate the role of value in the construction process of Alibaba's ecosystem to verify the effectiveness of the proposed value analysis framework. The relevant data comes from Alibaba's financial reports, official website and related service data in the APP Store [45,46].

With the development of the Internet economy, competition among enterprises is gradually surpassing the boundaries of individual enterprises and has evolved into competition among service ecosystems. The purpose of Internet companies to build their ecology is to realize a closed service loop through user sharing and resource sharing. In this way, users will become dependent on them, thereby increasing industry barriers. Alibaba is the most typical representative of Chinese Internet companies, and its service ecosystem has become increasingly mature.

Fig.10 shows the map of Alibaba's entire service ecosystem: e-commerce and financial services are core businesses; ports act as the role of user traffic guidance, such as UC.cn, aMap.com, and Weibo.com; local life services are current competition focus, including CaiNiao logistics, health care, and other offline businesses; some attempts are made in emerging areas, including gaming, video, music and other entertainment services. The core of this service ecosystem is data and traffic sharing, its foundation is marketing services and cloud services, and Alipay is the leader in the effective data integration.

(Small and Medium Enterprises) registered users with a free membership system. This has brought a steady information stream and created unlimited business opportunities, which has strengthened the online trading platform. In October 2003, Alipay was launched to solve the credit and security issues of online transactions. During this period, Alibaba continued to lay a solid foundation for its core business, and successively established and acquired Alimama, Koubei.com, Alibaba Software, etc.

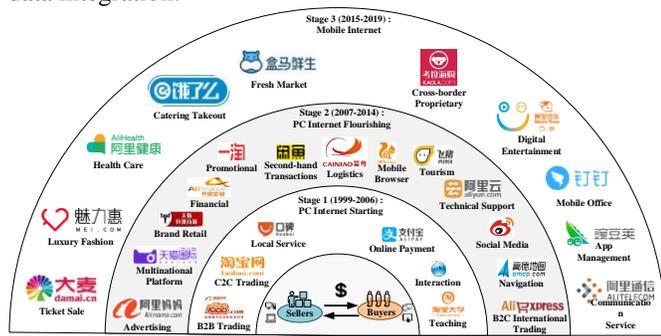

Fig.10 The evolution map of the Alibaba ecosystem

The evolution of Alibaba's service ecosystem is closely related to the value explosion caused by Internet technology. When the cost of Internet use is getting lower and lower, more and more industries and functions are penetrated by the Internet. Based on the time point of value explosion, we divided the evolution process of Alibaba's service ecosystem into three phases:

The first stage (1999-2006): During the initial stage of the Internet, retail business gradually went from offline to online. Alibaba used the "free + value-added fee" model to quickly occupy the market. Taobao attracted a large number of SME

The second stage (2007-2014): During this period, the Internet became popular and the number of online shopping users developed rapidly. Online retail transactions have gradually become an important part of people's lives. Alibaba realized an explosive growth in its core e-commerce business. In April 2008, Taobao entered the B2C field and launched vertical e-commerce businesses, such as Taobao Electric City, Taobao Famous Shoes Museum, and so on. In October 2010, Alibaba launched Yitao.com to build an independent shopping search engine for the entire Chinese e-commerce network.

The third stage (2015-present): At this stage, the mobile Internet began to rise, and smartphones replaced Personal Computers. Alibaba has expanded its investment scope to all walks of life. Relying on core e-commerce business, Alibaba made a lot of attempts in a number of areas, such as logistics services, cross-border e-commerce, local life services, etc. In its ecosystem, the overall synergy is constantly increasing: core business provides cash flow for other businesses; other fields provide support for core business through differentiated service provision.

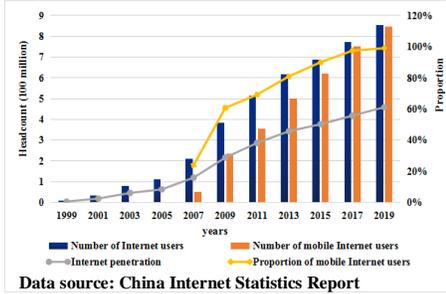
Data source: China Internet Statistics Report
（a）Statistics on the number and penetration rate of Internet users in China from 1999 to 2019

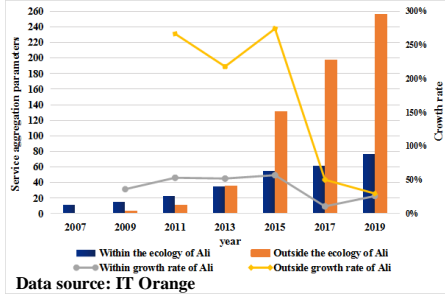
Data source: IT Orange
（b）The evolution of Alibaba's ecological scale in 2007-2019

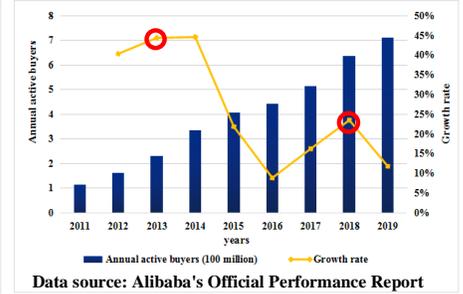
Data source: Alibaba's Official Performance Report
（c）Alibaba's annual number of active buyers and growth rate in 2011-2019

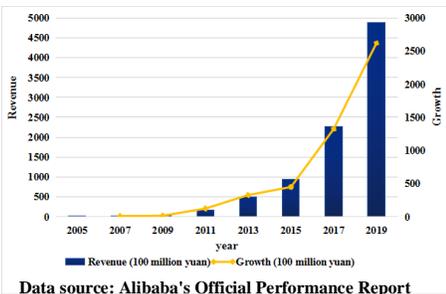
Data source: Alibaba's Official Performance Report
（d）Alibaba's annual revenue and growth in 2005-2019

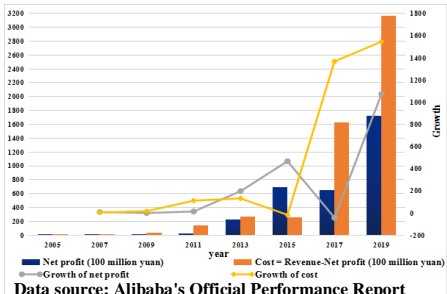
Data source: Alibaba's Official Performance Report
（e）Net profit and cost statistics of Alibaba Group in 2005-2019

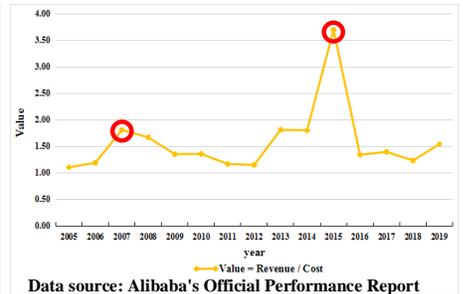
Data source: Alibaba's Official Performance Report
（f）Changes in the value of Alibaba Group in 2005-2019

Fig.11 Performance indicators of Alibaba ecosystem evolution

Fig.11 gives a quantitative analysis of the evolution of Alibaba's service ecosystem. The first row describes the evolution trend of the service ecosystem on the demand side, mainly the number of users. The second row describes the evolution trend of the service ecosystem on the supply side, mainly the trend of value changes.

Fig.11-a shows the number of Internet users in China from 1999 to 2019. It can be seen that the opportunity window for value explosion brought about by technological innovation. The first opportunity period was 2007, the number of Chinese Internet users increased rapidly from 210 million to 649 million. With the rapid growth of Internet users and the increasing popularity of the Internet, its commercial value has begun to shine. The second opportunity period was 2015, when the number of mobile Internet users has been close to the number of Internet users. The commercial value of the mobile Internet began to erupt in a short period of time.

Fig.11-b shows the population size of Alibaba's service ecosystem from 2007 to 2019. Here, the business services created and acquired by Alibaba itself are taken as its core services, and the business services invested by Alibaba are taken as its peripheral services. In 2015, the number of services in Alibaba's ecosystem began to explode (from 35 to 131) with the popularity of mobile payments. This is consistent with the changing trend of Fig.11-a.

Fig.11-c shows the trend of Alibaba's active users. In the early stage of development, the number of users has grown steadily. In 2013 and 2018, there were two peaks in the growth rate of its active users. According to the data analysis in Fig.11-b, we can know that Alibaba invested in Weibo in April 2013, and launched freshhema.com and ele.com in 2018. They have a very obvious user traffic guidance effect on the entire service ecosystem, which has directly caused an explosive growth in the number of users. This shows that the service convergence strategy has achieved a win-win effect in the Ali ecosystem.

Fig.11-d and Fig.11-e show Alibaba's revenue and costs from 2005 to 2019, respectively. According to formula 4, we set Alibaba's annual value as the ratio of revenue to cost. Fig.11-f shows the value curve of Alibaba. It can be seen that the value variable of Alibaba is always greater than 1.0, and two peaks appeared in 2007 and 2015 respectively. This is consistent with the results of the analysis of Fig.11-a, which correspond to two technological innovation opportunities: PC Internet popularity and mobile Internet explosion.

A closer examination of the evolution of Alibaba ecosystem reveals that value explosion is often accompanied by the maturity of technology innovation. At this stage, the output caused by emerging technologies will increase sharply, and operating costs will start to decrease sharply, such as the rapid increase in the number of users and services brought about by the popularity of mobile Internet. However, with a substantial growth in technology transfer and competitors, the output from emerging technologies is relatively stable, but operating costs have begun to increase. It indicates a possible shift from value explosion to value capturing as the market matures. This result thus supports the premise of our value analysis theory. It suggests that the emphasis of firms needs to move to explore the opportunity space to reduce cost through alliance formation at this stage.

## VI. CONCLUSION

In the service ecosystem, service providers, service consumers and service operators cooperate with each other to form a complex value network. The pursuit of value is the basic driving force for the evolution of service ecosystem. In order to promote the development of the entire service ecosystem, it is necessary to clarify its value-driven operation mechanism and design the appropriate intervention strategies. However, service ecosystem is a complex socio-technical system. Most of the related methods lack systematic research on its dynamic evolution mechanisms. .

In this context, we propose a value-driven analysis framework that can reveal the relationship between value network operations and service ecosystems evolution. The analysis framework includes four parts: value creation, value operation, value realization and value distribution. Experimental results and subsequent actual cases also prove the effectiveness of our proposed framework. The value analysis framework is universal because it does not depend on specific domain attributes. The above work can provide new research ideas and tools for the evolutionary analysis of service ecosystem.

The purpose of interpreting phenomena is to predict, while the purpose of prediction is to control. In the field of mobile Internet ecology, there are many trans-boundary cases where crossover services beat traditional services, such as Didichuxing.com vs traditional taxis, Internet finance vs traditional banks, mobile payments vs cash payments, etc. In the future, we will use the continuously optimized value model to analyze the trans-boundary phenomenon in the evolution of service ecosystems. Furthermore, we can reveal the explicit and implicit key factors affecting value, so as to provide the optimal evolution path of the service ecosystem in the specific context.

## REFERENCES

[1] Foster, I. Service-Oriented Science. Science, 2005,308(5723):814–817. doi:10.1126/science.1110411.

[2] Fox, A., Griffith, R., Joseph, A., Katz, R., Konwinski, A., Lee, G., Patterson,D., Rabkin,A., & Stoica, I.Above the clouds: A berkeley view of cloud computing. Dept. Electrical Eng. and Comput. Sciences, University of California, Berkeley, Rep. UCB/EECS, 2009,28(13) .

[3] Fuqing Yang , Lyu Jian, Mei Hong. Network structure software technology system: An Approach--Taking Architecture as Center. Science China (Evolume: Information Science), 2008(06): 818-828.

[4] Buera, F. J., & Kaboski, J. P. The rise of the service economy. American Economic Review, 2012,102(6):2540-2569.

[5] Allen, B., Bresnahan, J., Childers, L., Foster, I., Kandaswamy, G., Kettimuthu,R., Kordas, J., Link,M., Martin,S., Pickett,K., & Tuecke,S.Software as a service for data scientists. Communications of the ACM, 2012,55(2):81-88.

[6] Zhong, Y., Fan, Y., Huang, K., Tan, W., & Zhang, J.Time-aware service recommendation for mashup creation. IEEE Transactions on Services Computing, 2014,8(3): 356-368.

[7] Huang, K., Fan, Y., & Tan, W. Recommendation in an evolving service ecosystem based on network prediction. IEEE Transactions on Automation Science and Engineering, 2014,11(3): 906-920.

[8] Barros, A. P., & Dumas, M. The rise of web service ecosystems. IT professional,2006, 8(5):31-37.

[9] Xiao Xue, Yaodan Guo, Shufang Wang, Shizhan Chen. Analysis and Controlling of Manufacturing Service Ecosystem: A research Framework based on the Parallel System Theory. IEEE Transactions on Services Computing, DOI= 10.1109/TSC.2019.2917445, 2019.

[10] Huang, K., Liu, Y., Nepal, S., Fan, Y., Chen, S., & Tan, W.A novel equitable trustworthy mechanism for service recommendation in the evolving service ecosystem. In International Conference on Service-Oriented Computing 2014:510-517.

[11] Ghose, A., Ipeirotis, P. G., & Li, B.Examining the impact of ranking on consumer behavior and search engine revenue. Management Science, 2014, 60(7):1632-1654.

[12] West J, Wood D. Creating and Evolving an Open Innovation Ecosystem: Lessons from Symbian Ltd. Available at SSRN 1532926, 2008.

[13] Jansen, S., Handoyo, E., & Alves,C. Scientists' Needs in Software Ecosystem Modeling. Proceedings of the International Workshop on Software Ecosystems, 2015.

[14] Moore, J. F.Predators and prey: the new ecology of competition. Harward Business Review, 1993,71 (3):75-86.

[15] Lusch, R. F., & Vargo, S. L. Service-dominant logic: Premises, perspectives, possibilities. Cambridge University Press, 2014.

[16] Al-Masri, E., & Mahmoud, Q. H.Investigating web services on the world wide web.Proceedings of the 17th international conference on World Wide Web, 2008:795-804.

[17] Zheng, Z., Ma, H., Lyu, M. R., & King,I. Wsrec: A collaborative filtering based web service recommender system.IEEE International Conference on Web Services, 2009:437-444.

[18] Zheng, Z., Zhang, Y., & Lyu, M. R. Investigating QoS of real-world web services. IEEE transactions on services computing , 2012,7(1): 32-39.

[19] Cavallo, B., Di Penta, M., & Canfora, G. An empirical comparison of methods to support QoS-aware service selection. Proceedings of the 2nd International Workshop on Principles of Engineering Service-Oriented Systems ,2010: 64-70.

[20] Godse, M., Bellur, U., & Sonar, R. Automating qos based service selection.IEEE International Conference on Web Services , 2010:534-541.

[21] Villalba, C., & Zambonelli, F.Towards nature-inspired pervasive service ecosystems: Concepts and simulation experiences. Journal of Network and Computer Applications, 2011,34(2):589-602.

[22] Villalba, C., Rosi, A., Viroli, M., & Zambonelli, F.Nature-inspired spatial metaphors for pervasive service ecosystems. IEEE International Conference on Self-Adaptive and Self-Organizing Systems Workshops, 2008:332-337.

[23] Moustafa, A., Zhang, M., & Bai, Q. Trustworthy stigmergic service compositionand adaptation in decentralized environments. IEEE Transactions on services computing, 2014,9(2):317-329.

[24] Wang, H., Chen, X., Wu, Q., Yu, Q., Zheng, Z., & Bouguettaya, A.. Integrating on-policy reinforcement learning with multi-agent techniques for adaptive service composition. International Conference on Service-Oriented Computing , 2014: 154-168.

[25] Moustafa, A., Zhang, M., & Bai, Q. Trustworthy stigmergic service compositionand adaptation in decentralized environments. IEEE Transactions on services computing, 2014, 9(2):317-329.

[26] Moustafa, A., & Zhang, M. Multi-objective service composition using reinforcement learning. International Conference on Service-Oriented Computing, 2013:298-312.

[27] Wang, H., Wu, Q., Chen, X., Yu, Q., Zheng, Z., & Bouguettaya, A. . Adaptive and dynamic service composition via multi-agent reinforcement learning. IEEE International Conference on Web Services, 2014:447-454.

[28] Huang, G.Ma Y. , Liu X., LuoY. ,Lu, X., & Blake, B. .Model-Based Automated Navigation and Composition of Complex Service Mashups. IEEE Transactions on Services Computing, 2015,8(3):494-506.

[29] Tan, W., Zhang, J., Madduri, R., Foster, I., De Roure, D., & Goble, C. . Servicemap: Providing map and gps assistance to service composition in bioinformatics. IEEE International Conference on Services Computing , 2011:632-639.

[30] Fischer, R., Scholten, U., & Scholten, S..A reference architecture for feedback-based control of service ecosystems.IEEE International Conference on Digital Ecosystems and Technologies , 2010:1-6.

[31] Diao, Y. Using control theory to improve productivity of service systems. IEEE International Conference on Services Computing, 2007:435-442.

[32] Porter M E. Competitive advantage. New York: FreePress, 1985.

[33] Vargo S L, Lusch R F. Service-dominant logic: continuing the evolution. Journal of the Academy of marketing Science, 2008, 36(1): 1-10.

[34] Peppard J, Rylander A. From value chain to value network:: Insights for mobile operators. European management journal, 2006, 24(2-3): 128-141.

[35] Alves, C., de Oliveira, J. A. P., & Jansen, S. Software Ecosystems Governance-A Systematic Literature Review and Research Agenda. International Conference on Enterprise Information Systems, 2017:215-226.

[36] Haile N, Altmann J. Value creation in IT service platforms through two-sided network effects. International conference on grid economics and business models, 2012: 139-153.

[37] Touliou K, Bekiaris E. Building an Inclusive Ecosystem for Developers and Users: The Role of Value Propositions. Advances in Ergonomics Modeling, Usability & Special Populations, 2017: 339-346.

[38] Haile, N., & Altmann, J..Structural analysis of value creation in software service platforms. Electronic Markets, 2016, 26(2):129-142.

[39] Pant V, Yu E. Modeling strategic complementarity and synergistic value creation in coopetitive relationships.International Conference of Software Business, 2017: 82-98.

[40] Cong, P., Li, L., Zhou, J., Cao, K., Wei, T., Chen, M., & Hu, S. Developing user perceived value based pricing models for cloud markets.IEEE Transactions on Parallel and Distributed Systems, 2018,29(12):2742-2756.

[41] Duan, Y., Huang, K., Kattepur, A., & Du, W.Towards value-driven business modelling based on service brokerage.International Conference on Service-Oriented Computing 2013:163-176.

[42] Leng Liu, Ming Shen, Chengzhong Li. Challenges of application domain extension to value engineering theory. Journal of Beijing Institute of Technology, 01(2006):60-63.

[43] Panda, S., Modak, N. M., Basu, M., & Goyal, S. K. Channel coordination and profit distribution in a social responsible three-layer supply chain. International Journal of Production Economics, 168( 2015): 224-233.

[44] Xiao Xue, Shufang Wang, LejunZhang, Zhiyong Feng, Yaodan Guo. Social Learning Evolution (SLE): Computational Experiment-based Modeling Framework of Social Manufacturing. IEEE Transactions on Industrial Informatics, 2019, 15(6): 3343-3355.

[45] Alibaba's official annual performance report :https://www.alibabagroup.com/cn/ir/earnings.

[46] IT Orange：https://www.itjuzi.com.

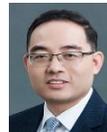
**Xiao Xue,** born in 1979. Professor with the School of Computer Software, College of Intelligence and Computing, Tianjin University. Also adjunct professor in the School of Computer Science and Technology, Henan Polytechnic University. His main research interests include service computing, computational experiment, complex network, etc.

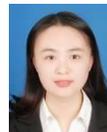
**Deyu Zhou,** born in 1995. Graduate student in the School of Information Engineering, China University of Geosciences (Beijing). Her current research interests include service computing and computational experiment.

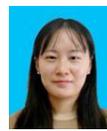
**Yaodan Guo**, born in 1994. Graduate student in the School of Computer Science and Technology, Henan Polytechnic University. Her current research interests include service computing.

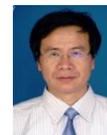
**Zhiyong Feng**, born in 1965. Professor and PHD supervisor in the School of Computer Software, Tianjin University. His main research interests include service computing, software engineering, Internet of things, etc

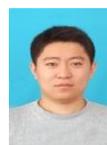
**Lejun Zhang,** born in 1979. Professor in the School of Computer Science and Technology, Yangzhou University. His research interests include computer network, social network analysis, dynamic network analysis and information security.

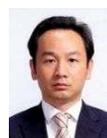
**Lin Meng,** born in 1978. Associate Professor with the Ritsumeikan University, Japan. His research interests include computer architecture, parallel processing, image recognition with Artificial Intelligence, IoT and so on. He is a member of IEICE, IEE, IPSJ and IEEE.